\documentclass{article}
\usepackage{amsmath}
\usepackage{graphicx}
\usepackage{dcolumn}
\usepackage{bm}
\usepackage{overpic}
\usepackage[T1]{fontenc}
\usepackage[utf8]{inputenc}
\usepackage{authblk}
\usepackage{booktabs}
\usepackage{color}
\usepackage{placeins}
\usepackage{geometry}
\usepackage[authoryear]{natbib}
\def\etal{\emph{et al.}}
\usepackage{amsmath,amsfonts,amssymb,bm,pifont}
\newcommand{\xmark}{\ding{55}}%
\usepackage[CaptionAfterwards]{fltpage}
\begin{document}
\title{The Time Dimension of Science: {Connecting the Past to the Future}}
\author[1,2]{Yian Yin}
\author[1,2,3]{Dashun Wang}
\affil[1]{Northwestern Institute on Complex Systems and Data Science (NICO), Northwestern University, Evanston, IL 60208, USA}
\affil[2]{McCormick School of Engineering and Applied Sciences, Northwestern University, Evanston, IL 60208, USA}
\affil[3]{Kellogg School of Management, Northwestern University, Evanston, IL 60208, USA}

\date{\today}
\maketitle

\begin{abstract}
A central question in science of science concerns how time affects citations. 
Despite the long-standing interests and its broad impact, 
we lack systematic answers to this simple yet fundamental question. 
By reviewing and classifying prior studies for the past 50 years, 
we find a significant lack of consensus in the literature, 
primarily due to the coexistence of retrospective and prospective approaches to measuring citation age distributions. 
These two approaches have been pursued in parallel, lacking any known connections between the two. 
Here we developed a new theoretical framework that not only 
allows us to connect the two approaches through precise mathematical relationships, 
it also helps us reconcile the interplay between temporal decay of citations and the growth of science, 
helping us uncover new functional forms characterizing citation age distributions. {We find retrospective distribution follows a lognormal distribution with exponential cutoff, while prospective distribution is governed by the interplay between a lognormal distribution and the growth in the number of references. Most interestingly, the two approaches can be connected once rescaled by the growth of publications and citations.}
We further validate our framework using both large-scale citation datasets and analytical models capturing citation dynamics.
Together this paper presents a comprehensive analysis of the time dimension of science, 
representing a new empirical and theoretical basis for all future studies in this area.

\end{abstract}

\section{Introduction}

The increasing availability of large-scale datasets that capture major activities in science has created an unprecedented opportunity to explore broad and important patterns underlying the scientific enterprise,
catalyzing in a drastic fashion a recent multidisciplinary shift in our quantitative understanding of science \citep{redner2005,guimera2005team, radicchi2008,jones2008multi,hirsch2005index,stringer2008effectiveness,newman2009first,petersen2014reputation,evans2008electronic, evans2011metaknowledge, szanto2014scientometrics,wang2013quantifying, uzzi2013atypical, sinatra2015century, ke2015defining, sinatra2016quantifying, guevara2016research}.
Nowhere are these new advances more apparent than in the studies of citations \citep{redner2005, bornmann2008citation,radicchi2008, eom2011characterizing,barabasi2012publishing,moreira2015distribution, wang2013quantifying, uzzi2013atypical}, owing largely to their widespread applications, from science policy to promotions \citep{lane2010let, radicchi2009diffusion,shen2014collective}, hiring \citep{clauset2015systematic, duch2012possible}, assignment of grants \citep{lane2011measuring,li2015big, bromham2016interdisciplinary} and prizes \citep{mazloumian2011citation}.
The aim of this paper is to carry out a comprehensive analysis on one of the most fundamental dimensions of citations: Time.

Time plays a {central} role in science.
Indeed, while some papers stay relevant and continue to dominate the scientific discourse over a long period, most papers unfortunately `die out', after collecting their fair share of citations \citep{wang2013quantifying,stringer2010}.
Understanding the time dimension is critical for a wide range of reasons.
It not only helps us understand the rise and fall of scientific paradigms \citep{kuhn2012structure},
tracing knowledge horizons {and hotspots} in science and technology \citep{sinatra2015century,orosz2016quantifying,mukherjee2017nearly}, {it is also critical for allocating investment \citep{szell2015research,ma2015anatomy, bromham2016interdisciplinary}}, identifying crucial yet initially unrecognized sleeping beauties \citep{van2004sleeping,ke2015defining}, assessing and even predicting future impact of inventions and discoveries \citep{wang2013quantifying,redner1998,uzzi2013atypical,acuna2012future,newman2014prediction, sinatra2016quantifying}.
This has become ever more so with the increasing scale, cost and complexity of science \citep{van2014top, borner2004simultaneous,vsubelj2016publication,michels2012growth}.
Indeed, the exponential growth of science \citep{price1963little,sinatra2015century,van2014top} suggests that there is now much more work for scientists to learn from, build upon, and cite,
which is further exacerbated by intensifying specialization in science and engineering disciplines \citep{jones2010nber}
and the inevitable dominance of interdisciplinary research spanning institutional and international boundaries \citep{jones2008multi,evans2009open,guimera2005team,deville2014career,adams2013nature}. Furthermore, because citation systems are commonly treated as models of complex interconnected systems, understanding the time dimension of citations will also help deepen our quantitative understanding of complex systems by tightening models and observations that are highly generalizable to broad areas.

\begin{FPfigure}

     \caption{{{\bf A} The first approach, denoted by \(P^{\leftarrow}(t)\), is also called `synchronous distribution'~ \citep{nakamoto1988synchronous}, `recent impact method'~ \citep{vinkler2010evaluation} or `citations from' approach as it measures the number of citations of a given age \emph{from} a paper~ \citep{redner2005}, or `retrospective citation' approach as it takes a paper and looks retrospectively back in time~ \citep{burrell2002modelling,glanzel2004towards}.
Denoted with \(P^{\rightarrow}(t)\), the second approach studies the age distribution of citations gained over time by a paper (or papers) published in a given year. Correspondingly, this method is also called `diachronous distribution'~ \citep{nakamoto1988synchronous}, `subsequent impact method'~ \citep{vinkler2010evaluation} or `citations to' approach~ \citep{redner2005}, or `prospective citation' approach~ \citep{burrell2002modelling,glanzel2004towards}. Here the width of arrows is proportional to the logarithm of citations, the size of the outer (inner) circle corresponds to the logarithm of total number of references (publications). {\bf B} Retrospective distribution for papers published in 2010, {lognormal with exponential cutoff (prediction by Eq (6), LN w. exp.) outperforms Weibull (W., normalized likelihood ratio $> 0~(599.3866),~p-value < 10^{-78016}$), lognormal (LN, normalized likelihood ratio $> 0~(169.2839),~p-value < 10^{-6225}$) and exponential (Exp., normalized likelihood ratio $>0~(173.0337),~p-value < 10^{-6503}$)}. {\bf C} Prospective distribution for papers published in 1950, {prediction by Eq (4) outperforms lognormal (LN, normalized likelihood ratio $>0~(50.0941),~p-value < 10^{-546}$) shifted power law (Shifted PL, normalized likelihood ratio $>0~(59.0745),~p-value < 10^{-759}$), power law with exponential distribution (PL w. exp., normalized likelihood ratio $>0~(202.7706),~p-value < 10^{-8930}$) and exponential (Exp., normalized likelihood ratio $>0~(96.9191),~p-value < 10^{-2041}$)}. {\bf D} Retrospective distribution for papers published in 1955, there are two dips due to world wars.} {\bf E} Average age of citations using retrospective and prospective approaches. Significant  fluctuations are observed for average age of citations until 1970. {\bf F} Retrospective distribution, $P^{\leftarrow}(t_2|t_1)$, in different years. {\bf G} Prospective distribution, $P^{\rightarrow}(t_1|t_2)$, in different years. Solid lines show a lognormal fit.}\label{overview}
  \includegraphics[width=1\textwidth]{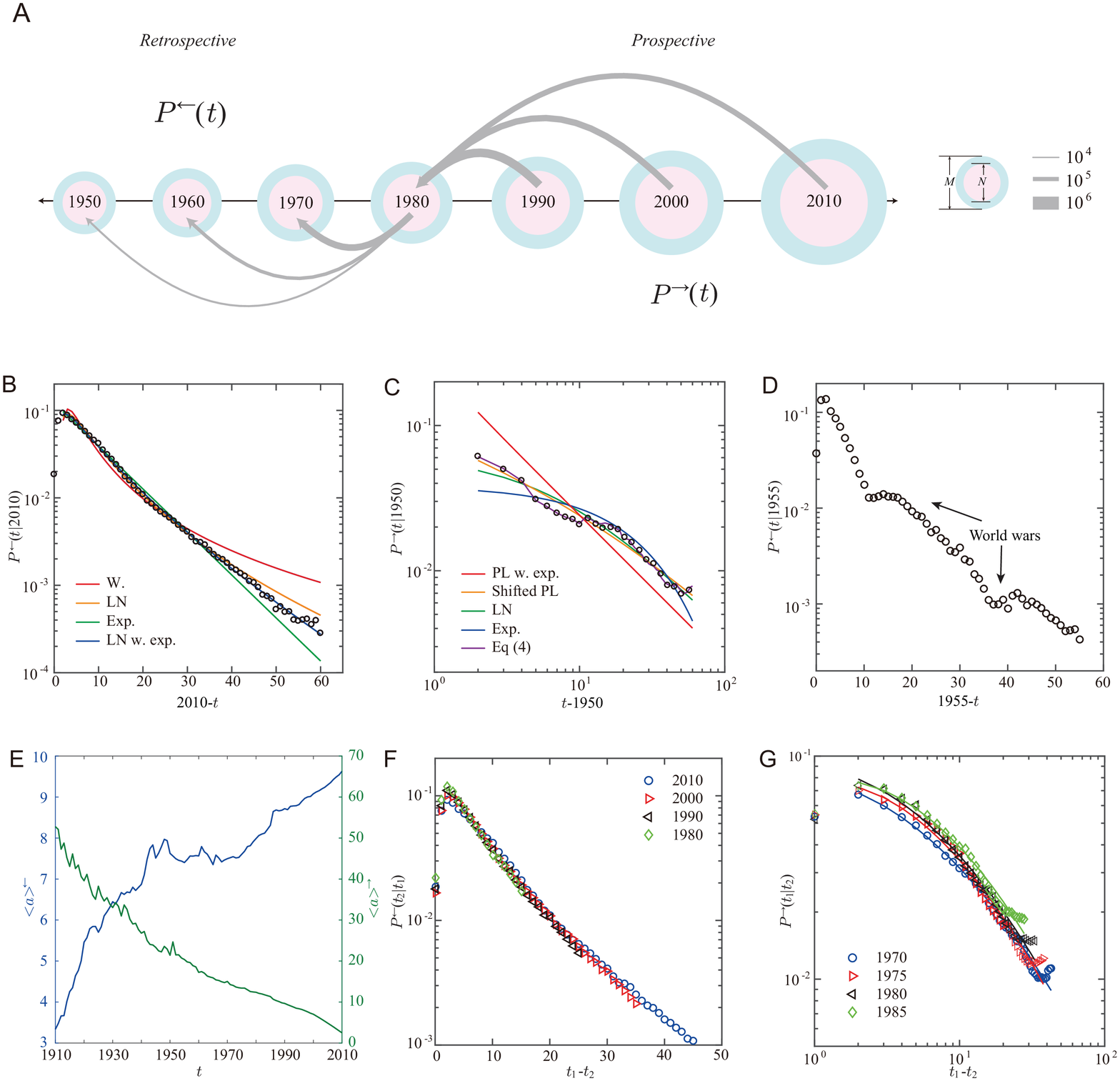}

\end{FPfigure}
\clearpage

It is, therefore, not surprising that this question {has been extensively investigated} over the past several decades \citep{redner2005,evans2008electronic,wang2013quantifying,borner2004simultaneous,burton1960half,garfield1963,price1965,margolis1967citation,macrae1969growth,krauze1971citations,avramescu1979actuality,stinson1987synchronous,nakamoto1988synchronous,motylev1989main,matricciani1991probability,egghe1992citation,gupta1997analysis,tsay1999library,pollmann2000forgetting,burrell2002modelling,glanzel2004towards,sanyal2006effect,simkin2007mathematical,lariviere2008long,bouabid2011revisiting,verstak2014shoulders,golosovsky2014uncovering,parolo2015attention,pan2016memory}, being one of most prolific lines of inquiry in science of science studies. Yet, despite its broad impact and rich historical context, we lack systematic answers to a simple yet fundamental question: How does time affect citations? In this paper, we systematically investigate this question by leveraging large-scale citation datasets and recent models capturing citations dynamics. We start by conducting a comprehensive review of existing literature, which reveals a significant lack of consensus on this matter. {The main reason of this lack of consensus} is that empirical measurements of temporal effect in citations have taken two related yet distinct measurement approaches (Fig. \ref{overview}A). The first approach considers papers that are cited by a publication and analyzes \textit{retrospectively} the age distribution of these references \citep{redner2005,burton1960half,garfield1963,price1965,margolis1967citation,macrae1969growth,krauze1971citations,avramescu1979actuality,stinson1987synchronous,nakamoto1988synchronous,motylev1989main,matricciani1991probability,egghe1992citation,gupta1997analysis,tsay1999library,pollmann2000forgetting,burrell2002modelling,glanzel2004towards,lariviere2008long,verstak2014shoulders,golosovsky2014uncovering,pan2016memory}. The second approach, in contrast, studies \textit{prospectively} the age distribution of citations that are gained over time by a paper \citep{redner2005,wang2013quantifying,macrae1969growth,krauze1971citations,avramescu1979actuality,stinson1987synchronous,nakamoto1988synchronous,motylev1989main,glanzel2004towards,sanyal2006effect,simkin2007mathematical,bouabid2011revisiting,golosovsky2014uncovering,parolo2015attention}.
The subtle difference between the two approaches creates a dramatic yet largely understudied effect in temporal citation patterns (Table \ref{tab:age}), further confounded by the exponential growth in both the number of papers and references cited by them \citep{sinatra2015century,van2014top,pan2016memory}.\par

\begin{table}[!]
\centering
\renewcommand\arraystretch{0.9}
\begin{tabular}{l|c|c|c|c}
\hline
Paper & \(P^{\leftarrow}(t)\) &  \(P^{\rightarrow}(t)\) &  \(P^{\leftarrow}_{N}(t)\) &  \(P^{\rightarrow}_{r}(t)\)\\
  \hline
 \cite{burton1960half} & Exp. v3 & \xmark & \xmark & \xmark \\
 \cite{garfield1963} & --- & \xmark & \xmark & \xmark \\
\cite{price1965} & Exp. & \xmark  & Const. & \xmark \\
 \cite{margolis1967citation} & Exp. & \xmark & \xmark & \xmark \\
\cite{macrae1969growth} & Exp. & Exp. & Exp. & \xmark\\
 \cite{krauze1971citations} & Exp. & Exp. & Exp. &\xmark\\
 \cite{avramescu1979actuality} & Exp. v2 & --- & \xmark. & \xmark \\
 \cite{stinson1987synchronous} &--- &---&---&---\\
 \cite{nakamoto1988synchronous} &Exp. & Exp.& \xmark & \xmark \\
\cite{motylev1989main} & Exp. v3&--- &--- &\xmark\\
 \cite{matricciani1991probability} &LN &\xmark &\xmark&\xmark\\
 \cite{egghe1992citation} & LN\(>\)W., Exp. v2, NB & \xmark & \xmark & \xmark \\
\cite{gupta1997analysis} & LN & \xmark & \xmark & \xmark \\
\cite{tsay1999library} & Exp. &\xmark &\xmark &\xmark\\
 \cite{pollmann2000forgetting} & Exp. v1\(>\) Exp. & \xmark  &\xmark &\xmark\\
\cite{burrell2002modelling} & LN\(>\)LL, W. & \xmark & \xmark & \xmark \\
 \cite{glanzel2004towards} &--- & PL &\xmark &\xmark \\
\cite{borner2004simultaneous} &W.&\xmark&\xmark&\xmark\\
\cite{redner2005} & Exp. & PL & PL & --- \\
\cite{sanyal2006effect} & \xmark & LN & \xmark & \xmark \\
 \cite{simkin2007mathematical}&\xmark & PL w. Exp.&\xmark &\xmark\\
\cite{lariviere2008long} & --- & \xmark & \xmark &\xmark\\
 \cite{evans2008electronic} & --- & \xmark & \xmark & \xmark\\
 \cite{bouabid2011revisiting} &\xmark & LL&\xmark& \xmark\\
\cite{wang2013quantifying} &\xmark & LN&\xmark &\xmark\\
\cite{verstak2014shoulders} &--- &\xmark &\xmark &\xmark\\
 \cite{golosovsky2014uncovering} &--- &---&\xmark &\xmark\\
\cite{parolo2015attention} &\xmark & Exp. v1\(>\)PL w. const&\xmark &\xmark\\
\cite{pan2016memory} &Exp. &--- &---&\xmark\\
\hline
\end{tabular}

\caption{ {Age distribution of citations: retrospective vs.~prospective approaches. Not measured: \xmark. Measured but no functional form was offered: ---. See Table \ref{tab:damping} for explanations of abbreviations. \(X>Y\): \(X\) outperforms \(Y\) during fitting for \(P(t)\)}}
\label{tab:age}
\end{table}

\begin{table}[!]
\renewcommand\arraystretch{0.9}
\begin{center}

\begin{tabular}{c|c|c}
\hline
{Abbreviation} &{Full Name} & {{Functional Form} of \(P(t)\)}\\
  \hline
Const. &Constant& \large{\(1\)}\\
Exp. &Exponential& \large{\(e^{-\lambda t}\)}\\
Exp. v1 &Exponential variant 1& \large{\(\beta e^{-\lambda t}+\gamma\)}\\
Exp. v2 &Exponential variant 2& \large{\(e^{-\lambda t}-e^{-m\lambda t}\)}\\
Exp. v3 &Exponential variant 3&\large{\(ae^{-x}+2(1-a)e^{-2x}\)}\\
PL &Power law& \large{\(t^{-\alpha}\)}\\
PL. w. const &Power law with constant& \large{\(\beta t^{-\alpha}+\gamma\)}\\
Pow. w.~Exp. &Power law with exponential cutoff & \large{\(t^{-\alpha}e^{-\lambda t}\)}\\
W. &Weibull&\large{\((\frac{t}{\alpha})^{\beta-1}e^{-(\frac{t}{\alpha})^{\beta}}\)}\\
NB&Negative Binomial&\large{\(\frac{\Gamma(k+t)}{\Gamma(k)\Gamma(t+1)}p^kq^t\)}\\
LN &Lognormal& \large{\(\frac{1}{t}e^{-\frac{(\ln(t)-\mu)^2}{2\sigma^2}}\)}\\
LL &Log-logistic& \large{\(\frac{(x/\alpha)^{\beta-1}}{(1+(x/\alpha)^{\beta})^2}\)}\\
\hline
\end{tabular}

\caption{{Distributions appeared in Table \ref{tab:age} }\label{tab:damping}}

\end{center}
\end{table}

The temporal behavior of citations has been measured and reported by independent research {groups, each using specific datasets} and measurement details (Table \ref{tab:age}). The lack of consensus, and the coexistence of the two approaches, raise an important question: What is the most appropriate functional form describing the time dimension of science?
Here, by introducing a general theoretical framework, we provide systematic answers to this question, which is then validated both analytically and empirically through citation models and large-scale datasets.
As such, the paper makes several contributions to this prolific direction in the science of science.
First and foremost, it serves as a detailed and much needed survey by reviewing and repeating results from most major studies in this domain. We then used the Web of Science (WoS) dataset to systematically measure and test results obtained by prior studies.
We examine these results in the context of growth in science, systematically testing the most appropriate model describing aging measured by the prevailing two approaches. More importantly, despite the concomitant development of both approaches in the literature, we lack any known connections between the two.
Here we introduced a general framework that helps us uncover that the two different approaches are connected through precise mathematical relationships, allowing us to derive and even predict one approach from the other.
We further derived our empirical results using citation models, providing theoretical support for the observed temporal behavior.
We conclude by a brief discussion on scaling relationships between aging structure and citation impact of papers.
Together, this paper presents a comprehensive study on the time dimension of science, providing a new empirical and theoretical basis for all further studies on this topic.
These results are not only relevant for the emerging {field of the science of science}, they also help improve our understanding of complex systems, by pushing forward measurements and models in complex evolving networks.

\section{Results}

{A frequently applied measure for the time dimension of science is the age distribution of citations}. Since 1960s, there have been numerous studies reporting on this distribution by taking one or both of the two approaches. While both approaches focus on the age of citations, they measure quantities pertaining to rather different processes (Fig.~\ref{overview}A). Indeed, the retrospective approach, denoted by $P^{\leftarrow}(t)$, measures how far back the paper looks in time to cite old references, hence quantifying the citing memory of a paper.
The prospective approach $P^{\rightarrow}(t)$, on the other hand, measures how a paper is cited by others in the future, effectively measuring how the research community {remembers} a paper over time.
Here, we review all major studies on this topic and summarize them in Table \ref{tab:age}, asking specifically: what approach did they take? And what results did they obtain?
Most early studies have been focused on the retrospective approach, a tradition dating back at least to Burton and Kebler \citep{burton1960half}. Out of the seventeen papers that reported explicitly the functional form of $P^{\leftarrow}(t)$ in Table \ref{tab:age}, twelve of them supported the exponential model or its variants to be the best fit to data.
Since MacRae compared the retrospective and prospective citation age distributions in 1969 \citep{macrae1969growth}, more and more studies have started to focus on the prospective approach to measuring age distribution of citations. Comparing the two columns under \(P^{\leftarrow}(t)\) and \(P^{\rightarrow}(t)\) in Table \ref{tab:age} {reveals that the prospective distributions are typically characterized by broader distributions than retrospective distributions}, ranging from power law to power law with an exponential cutoff to lognormal distributions.
Note that the different results reported in Table~\ref{tab:age} do not necessarily imply that these papers contradict with each other, because they strongly depend on measurement details as well as the corpus of papers on which these studies were carried out. The remarkably diverse conclusions summarized in Table \ref{tab:age}, however, raise a tantalizing question: When it comes to citation age distributions, for both \(P^{\leftarrow}(t)\) and \(P^{\rightarrow}(t)\), what are the most appropriate functional forms that capture their temporal behavior? To answer this question, we take the Web of Science dataset, arguably the most authoritative citation index, to systematically measure both retrospective and prospective distributions.\par
\subsection{Empirical Measurements}
To understand \(P^{\leftarrow}(t)\), we first test the three most frequently {applied distributions:} Weibull, exponential and lognormal distributions (Table \ref{tab:damping}), asking which one best fits our data. We use maximum likelihood estimation  in this paper to determine the fit (Sec. S1). Since the tail part of the distribution is of the most interest {and the citations in the first year after publication can be largely affected by publication date}, we follow recent approaches by Parolo \etal ~\citep{parolo2015attention} and focus on data after the peak for the fit (typically occurring around 2 years).
More specifically, We take all papers published in the same year (e.g. year 2010), trace retrospectively back in time, and measure the number of citations from the focal year to each earlier time unit. We find among the three functions tested, lognormal distribution provides the best fit for $P^{\leftarrow}(t)$ in terms of log-likelihood estimation, outperforming both Weibull and exponential distributions (Fig.~\ref{overview}B). However, none of them provides a satisfactory fit, showing an evident deviation for the tail part, hinting for alternative models that can fit \(P^{\leftarrow}(t)\) more accurately. Indeed, we use a lognormal distribution with exponential cutoff, for reasons that will be explained later, and find it provides a better fit to the data (outperforming other distributions from likelihood ratio test with $p-value<10^{-6225}$).

{For the prospective distribution} $P^{\rightarrow}(t)$, we tested four types of functions reported in Table \ref{tab:age}, from exponential distribution to lognormal to power law with and without exponential cutoff (Table \ref{tab:damping}). We also used shifted power law, i.e. $P(t)\sim (t_0+t)^{\alpha}$, to fit our data. We calculated \(P^{\rightarrow}(t)\) using the same set of papers as Figure \ref{overview}B, finding that lognormal function outperforms other distributions (Fig. \ref{overview}C). Interestingly, power law with and without exponential cutoff yield similar results, because the obtained exponential cutoff approaches its lower bound, playing a minor role in distinguishing the two distributions. {Indeed, as we show later, the prospective distribution follows Eq (4), which not only gives a better fitting (outperforming other distributions from likelihood ratio test with $p-value<10^{-546}$), but also captures all the fluctuations in empirical data.}

Both distributions can be affected by significant external shocks to the system. Indeed, repeating our analysis for year 1955 reveals two `dips' in the retrospective distribution around 1942 and 1919, respectively (Fig. \ref{overview}D), documenting interruptions created by wars. To quantify the extent to which the system is affected by external forces, we calculate the average age of all citations both from and to papers published at time $t$, denoted by $\langle a\rangle^{\leftarrow}(t)$ and $\langle a\rangle^{\rightarrow}(t)$, respectively. Figure \ref{overview}E shows that $\langle a\rangle^{\leftarrow}(t)$ shows larger fluctuations prior to 1970. Hence in this paper, we measure $P^{\leftarrow}$ for data from 1970 to 2013. For $P^{\rightarrow}$, we use data after the {Second World War}, i.e. from 1948 to 2013.

Figure \ref{overview}E also demonstrates that the aging structure of citations changes over time \citep{sinatra2015century,verstak2014shoulders,pan2016memory}. To further understand temporal evolution of the aging structures, we calculate \(P^{\leftarrow}(t)\) and $P^{\rightarrow}(t)$ for different years \(t_0\). If we group paper samples based on their publication year $t_0$, mathematically this can be described as \(P^{\leftarrow}(t|t_0)\) and $P^{\rightarrow}(t|t_0)$. To this end we assume $t_1$ is a more recent year than $t_2$ ($t_1>t_2$). Then $P^{\leftarrow}(t_2|t_1)$ measures the probability that a citation from $t_1$ points to a reference published at time $t_2$. Similarly $P^{\rightarrow}(t_1|t_2)$ measures the probability that a paper published in $t_1$ is later cited at $t_2$. We plot both distributions for different years (Fig.~\ref{overview}F,G). For $P^{\leftarrow}(t_2|t_1)$, we find papers published in different years after 1980s follow a similar citation pattern, with recent papers having a stronger tendency to cite `old papers' (Sec. S2), which supports earlier findings by using Google Scholar dataset \citep{verstak2014shoulders}. Comparing $P^{\leftarrow}(t_2|t_1)$ (Fig. \ref{overview}F) with average reference ages for papers published in different years (Fig.~\ref{overview}E), we find this increasing preference for old papers can be traced back to the 1970s, after leaving an unstable regime in the 1960s. 

Most important, we find in many cases $P^{\rightarrow}(t)$ is not a monotonically decreasing function. Indeed, we observed ubiquitous `upward tails' characterizing temporal behavior of citations.  For example, papers published in 1970 received 7.87\% more citations in 2013 than they did in 2006. This indicates, as papers get older, they collect more instead of fewer citations. Repeating this analysis for different years, we find the `upward tails' characterizing \(P^{\rightarrow}(t_1|t_2)\) is nearly universal (Fig. \ref{overview}G).  
The upward tails offer the strongest demonstration that none of the functions we have discussed so far (Table \ref{tab:age}) are adequate in describing $P^{\rightarrow}(t)$. This phenomenon has been reported in prior works \citep{sinatra2015century}, but has received scarce attention in the literature.  
Next we show, understanding these tails not only helps us understand the role of growth in science in driving age distributions of citations, but also enables us to uncover the correct functional forms describing this temporal behavior.
\subsection{Accounting for Growth}
\begin{figure}[!]
\centering
\includegraphics[width=0.667\linewidth]{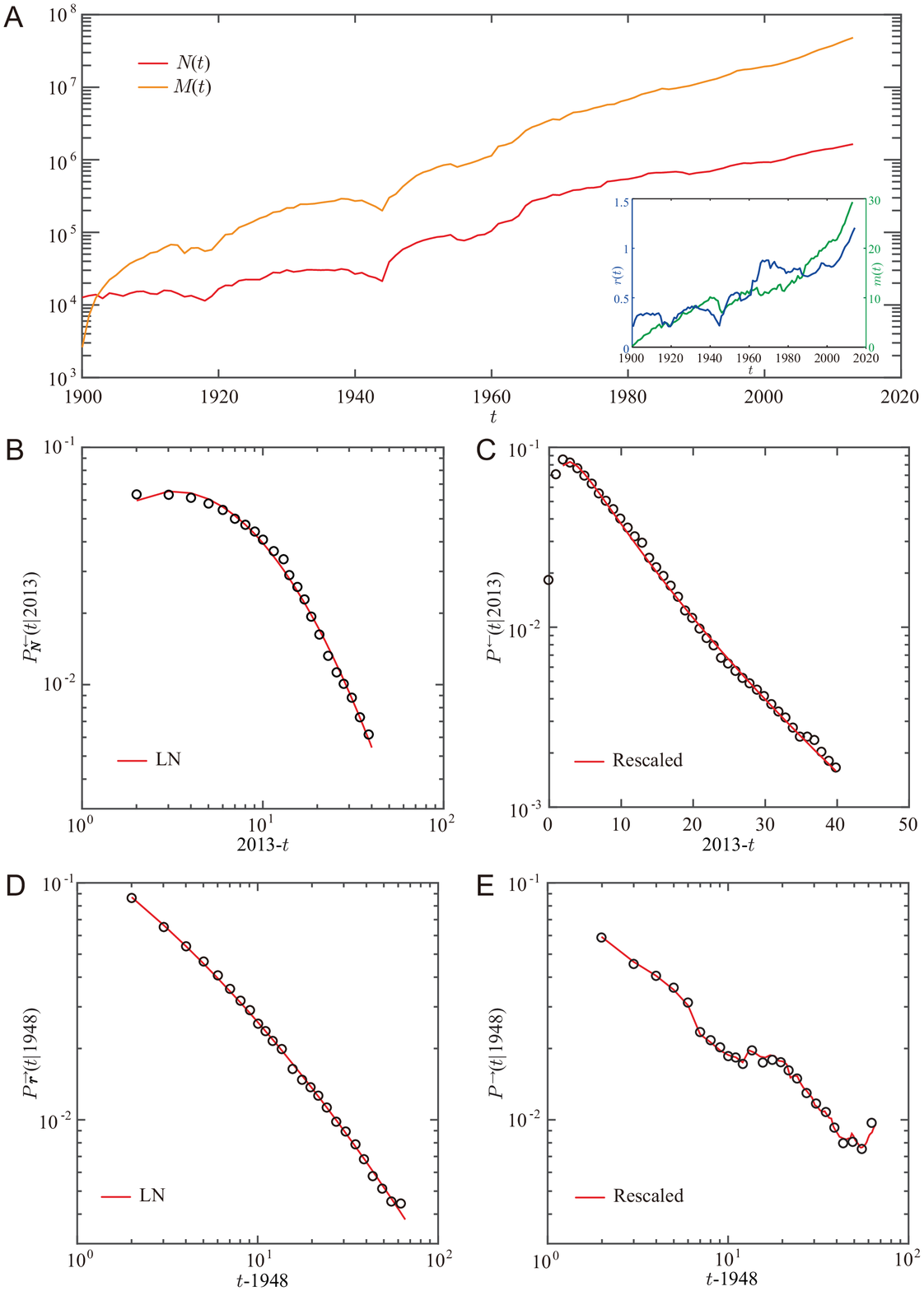}
\caption{{\bf A} Number of papers published and citations generated in every year grows exponentially. {Inset}: The growth in the average number of references from papers published every year. {\bf B} Retrospective distribution normalized by \(N\) in 2013, with lognormal showing the best fit. {\bf C} Retrospective distribution in 2013. The fitting is recovered from normalized distribution, i.e. lognormal with exponential cutoff is the best fit. {\bf D} Prospective distribution normalized by \(r\) for papers published in 1948, with lognormal showing the best fit. {\bf E} Prospective distribution in 1948, the fitting is recovered from normalized distribution. The best fit reproduces the upward tail.}
\label{normalize}
\end{figure}

\(P^{\leftarrow}(t)\) and \(P^{\rightarrow}(t)\) are determined by both aging of citations and growth of the citation system. Indeed, if we look further back in time, there are inherently fewer papers to cite, hence affecting \(P^{\leftarrow}(t)\). Similarly, the continuous expansion of science impacts how an existing paper is remembered over time as new citations  enter the system at an increasing rate. Let us denote with \(N(t)\), the number of papers published at time \(t\). Extensive literatures documented that \(N(t)\) is best approximated by an exponential function: $N(t) \sim e^{\beta t}$ \citep{wang2013quantifying,sinatra2015century,van2014top} (Fig.~\ref{normalize}A). Besides the sheer increase in the number of papers, the average number of references contained in each paper, \(m(t)\), also increases with time (Fig. \ref{normalize}A inset). Indeed, if we define \(M(t)\) as the number of citations from papers published at time \(t\), i.e., \(M(t)\equiv N(t)m(t)\), to the leading order the growth of \(M\) can also be approximated by an exponential function (Fig. \ref{normalize}A). To account for the growth of science, we need to normalize the obtained retrospective distributions by the number of papers available for citations

\begin{equation}
P_{N}^{\leftarrow}(t_2|t_1)=\frac{P^{\leftarrow}(t_2|t_1)/N(t_2)}{\int_{0}^{t_1}P^{\leftarrow}(t_2|t_1)/N(t_2)dt_2},
\label{eq:rretro}
\end{equation}
where \(P^{\leftarrow}_{N}(t_2|t_1)\) captures the fact that \(P^{\leftarrow}(t_2|t_1)\) is normalized by the number of papers available for reference at time $t_2$, \(N(t_2)\).
\par
\(M(t_1)\) new citations point towards \(\int_{0}^{t_1}N(t)dt\) papers in the system. Thus, for the prospective distribution, we normalize \(P^{\rightarrow}(t)\) by the likelihood to be cited by new references, \(r(t_1)\equiv\frac{M(t_1)}{\int_{0}^{t_1}N(t)dt}\), yielding
 \begin{equation}
P_{r}^{\rightarrow}(t_1|t_2)=\frac{P^{\rightarrow}(t_1|t_2)/r(t_1)}{\int_{t_2}^{T}P^{\rightarrow}(t_1|t_2)/r(t_1)dt_1}
\label{eq:rpro}
\end{equation}
Note that $r(t_1)$ reduces to \(\beta m(t_1)\) in the condition of \(N(t)\sim e^{\beta t}\) (Fig.~\ref{normalize}A inset). Here we use \(N(t)\) to normalize retrospective distribution, which is equivalent to a null model that assumes every paper has the same chance to be cited. The prospective distribution, on the other hand, is normalized by \(r(t)\), indicating the probability for a paper to be cited at a given time is driven by both growth of papers and references. Only a modest amount of literature measured temporal effect on citations by taking into account growth factors (Table \ref{tab:age}).
To find the best functional form for $P^{\rightarrow}_{r}(t_1|t_2)$, we take all papers published in a year, measure the number of citations they receive in subsequent years, and normalize the distributions by \(r(t)\). For \(P^{\leftarrow}_{N}(t)\), we select papers from year 2013, and normalize their \(P^{\leftarrow}(t)\) by $N(t)$. Although \(P^{\rightarrow}(t)\) is characterized by a broader distribution than \(P^{\leftarrow}(t)\), once normalized however, we find both distributions are best fitted by the same function: lognormal distribution  (Fig.~\ref{normalize}BD). That is,
\begin{equation}
P_N^{\leftarrow}(t_2|t_1)\sim \frac{1}{\sqrt{2\pi}\sigma(t_1-t_2)}e^{-(\ln(t_1-t_2)-\mu)^2/2\sigma^2}
\label{eq:res1}
\end{equation}
\begin{equation}
P_r^{\rightarrow}(t_2|t_1)\sim \frac{1}{\sqrt{2\pi}\sigma(t_1-t_2)}e^{-(\ln(t_1-t_2)-\mu)^2/2\sigma^2}
\label{eq:res2}
\end{equation}\par
Indeed, \(P^{\leftarrow}(t)\) is simultaneously driven by two factors: the obsolescence of citations, $P_{N}^{\leftarrow}(t)$, and the growth of science, $N(t)$. Since the latter leads to an exponential decay when we look backwards, the true aging function can be hidden from the rapid growth of science when measuring \(P^{\leftarrow}\) directly. To this end, we use (\ref{eq:rretro}) and (\ref{eq:res1}) to recover the functional form of \(P^{\leftarrow}\), obtaining
\begin{equation}
P^{\leftarrow}(t_2|t_1)\sim \frac{1}{\sqrt{2\pi}\sigma(t_1-t_2)}e^{-(\ln(t_1-t_2)-\mu)^2/2\sigma^2-\beta(t_1-t_2)}
\label{eq:pretro}
\end{equation}
where \(\mu\) and \(\sigma\) are estimated from \(P_{N}^{\leftarrow}\). Note that \(\beta\) characterizes the exponential growth of publications, which is a fixed system-wide parameter. Therefore, (\ref{eq:pretro}) has the same number of independent parameters as other fitting functions analyzed in this paper. Yet, when fitted to retrospective distributions, (\ref{eq:pretro}) offers a superior fit than all functions discussed in Table \ref{tab:age} (Fig. \ref{normalize}C). Furthermore, the exponential part dominates over the lognormal factor in the long term, explaining why it appears to be approximated by an exponential function in most earlier cases.\par
Similarly, we recover the mathematical form of \(P^{\rightarrow}\) from \(P_{r}^{\rightarrow}\), obtaining
\begin{equation}
P^{\rightarrow}(t_1|t_2)\sim \frac{r(t_1)}{\sqrt{2\pi}\sigma(t_1-t_2)}e^{-(\ln(t_1-t_2)-\mu)^2/2\sigma^2}
\label{eq:ppro}
\end{equation}
(\ref{eq:ppro}) not only provides an excellent fit to empirical data (Fig. \ref{normalize}E), it also predicts accurately the emergence of upward tails: since $r(t_1)$ increases with time, it competes with aging of citations, compensating the decay in (in some case even increasing)  $P^{\rightarrow}(t)$. Using empirical data of $r(t)$, we successfully reproduce these tails, documenting the validity of (\ref{eq:ppro}) (Fig. \ref{normalize}E). Note that when $t_1-t_2$ is small, $\frac{\partial\ln(P^{\rightarrow}(t_1|t_2))}{\partial t_1}$ can be approximated as $\frac{\ln(t_1-t_2)-\mu+\sigma^2}{\sigma^2(t_1-t_2)}$, explaining why lognormal appears to fit $P^{\rightarrow}$ well when $t_1-t_2$ is small. Moreover, (\ref{eq:ppro}) indicates that, for growth factor to dominate over the aging factor, $t_1-t_2$ needs to be large enough, explaining why upward tails are only evident for old papers. 

Taken together, our results show that the time dimension of science is different from any result reported in Table \ref{tab:age}: First of all, both distributions follow lognormal distributions after appropriate normalization. $P^{\leftarrow}(t)$ can be modeled by a lognormal distribution with exponential cutoff, where the cutoff is caused by the exponential growth of papers. $P^{\rightarrow}(t)$ is the product of a lognormal function and $r(t)$, which lacks clear mathematical forms. A recent literature \citep{pan2016memory} suggests $r(t)$ to be an exponential function, which means $P^{\rightarrow}(t_1|t_2)$ will be dominated by the growth of $r(t)$ and increases infinitely as $t_1$ increases. However, if we assume $r(t)$ to grow with a slower rate, e.g. following a linear or power law function \citep{ucar2014growth,krapivsky2005network,vazquez2001statistics,allen1994persuasive}, the upward tail is only temporary and $P^{\rightarrow}(t_1|t_2)$ will be eventually dominated by the lognormal decay from $P_{r}^{\rightarrow}(t)$.

{To test the robustness of (\ref{eq:pretro}) and (\ref{eq:ppro}), we further repeat our empirical measurement for thirty years (See Fig. S2,3 in Supplementary Information), finding a universal pattern of citation decay across decades. Next we show, (\ref{eq:pretro}) and (\ref{eq:ppro}) not only offer more accurate mathematical forms to fit the distributions, they also allow us to build a conceptually new framework which is closely related to several citation models.}

\subsection{Validations}

Given the new results obtained in (\ref{eq:pretro}) and (\ref{eq:ppro}), we must ask: how do we validate these results? Next we show two types of validations, by deriving both empirical and analytical evidence that supports the validity of (\ref{eq:pretro}) and (\ref{eq:ppro}).

\begin{figure}[h]
\centering
\includegraphics[width=0.667\linewidth]{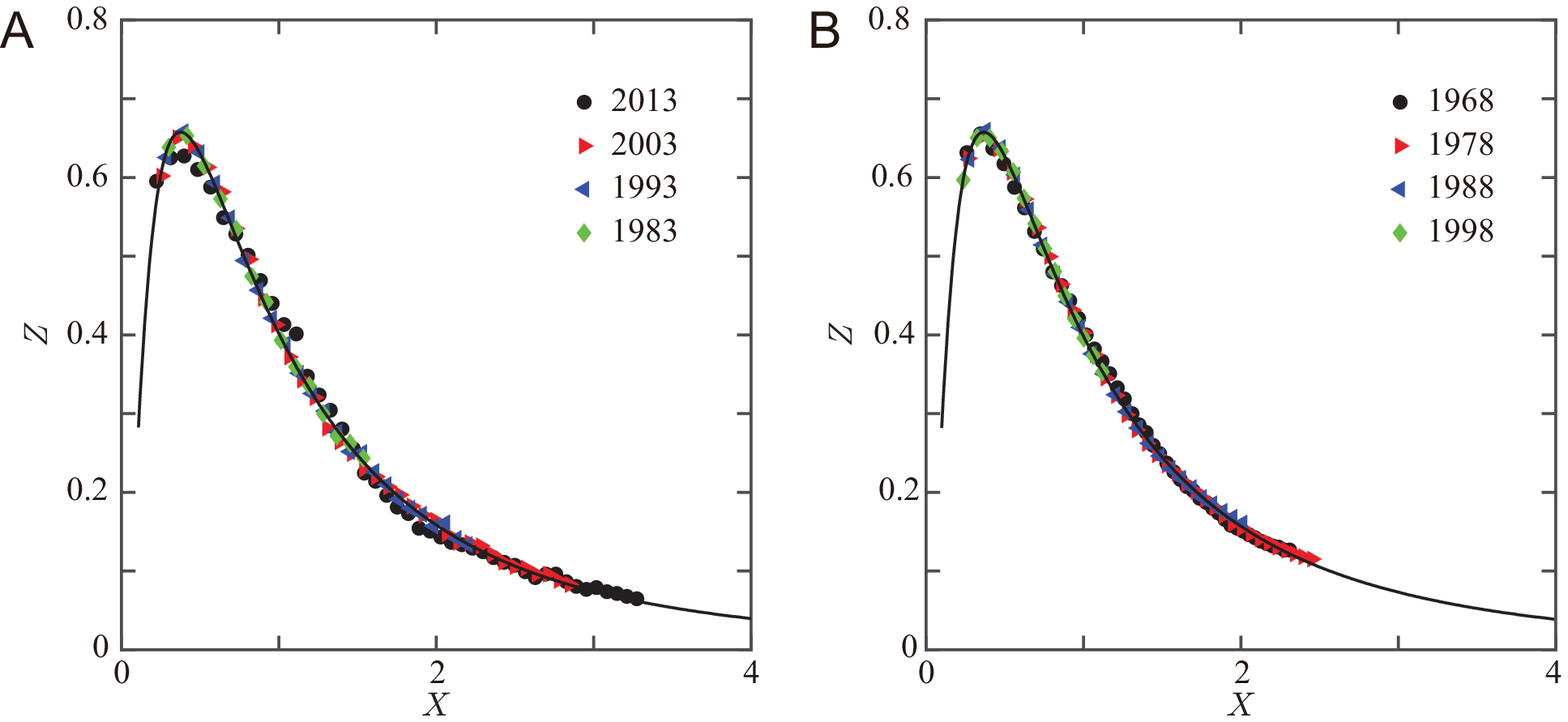}
\caption{{\bf Empirical support for $P^{\leftarrow}(t)$ and $P^{\rightarrow}(t)$.} {\bf A} \(P^{\leftarrow}_{N}(t)\) rescaled with respect to \(\mu\) and \(\sigma\). Distributions from different years collapse on the probability density function (PDF) curve of standard lognormal distribution. {\bf B} \(P^{\rightarrow}_{r}(t)\) rescaled with respect to \(\mu\) and \(\sigma\). Similar to A, distributions from different years collapse on the PDF curve of standard lognormal distribution.}
\label{new_evidence}
\end{figure}

To validate (\ref{eq:pretro}) empirically, we use normalized retrospective distribution from different years and rescaled them with respect to \(\mu\) and \(\sigma\). 
(\ref{eq:pretro}) and (\ref{eq:ppro}) predicts that both distributions collapse into standard lognormal distribution, i.e. \(Z=\frac{1}{\sqrt{2\pi}X}e^{-(\ln X)^2/2}\), after a proper rescaling (Sec. S3). We test this hypothesis empirically for $P^{\leftarrow}_{N}$, finding excellent data collapse across different years (Fig. \ref{new_evidence}A). We repeat this test for \(P_{r}^{\rightarrow}\) using (\ref{eq:ppro}), finding again an excellent data collapse on the same curve (Fig. \ref{new_evidence}B). The data collapses documented in Figure \ref{new_evidence} offer strong empirical evidence for (\ref{eq:pretro}) and (\ref{eq:ppro}).

\begin{figure}[h]
\centering
\includegraphics[width=0.667\linewidth]{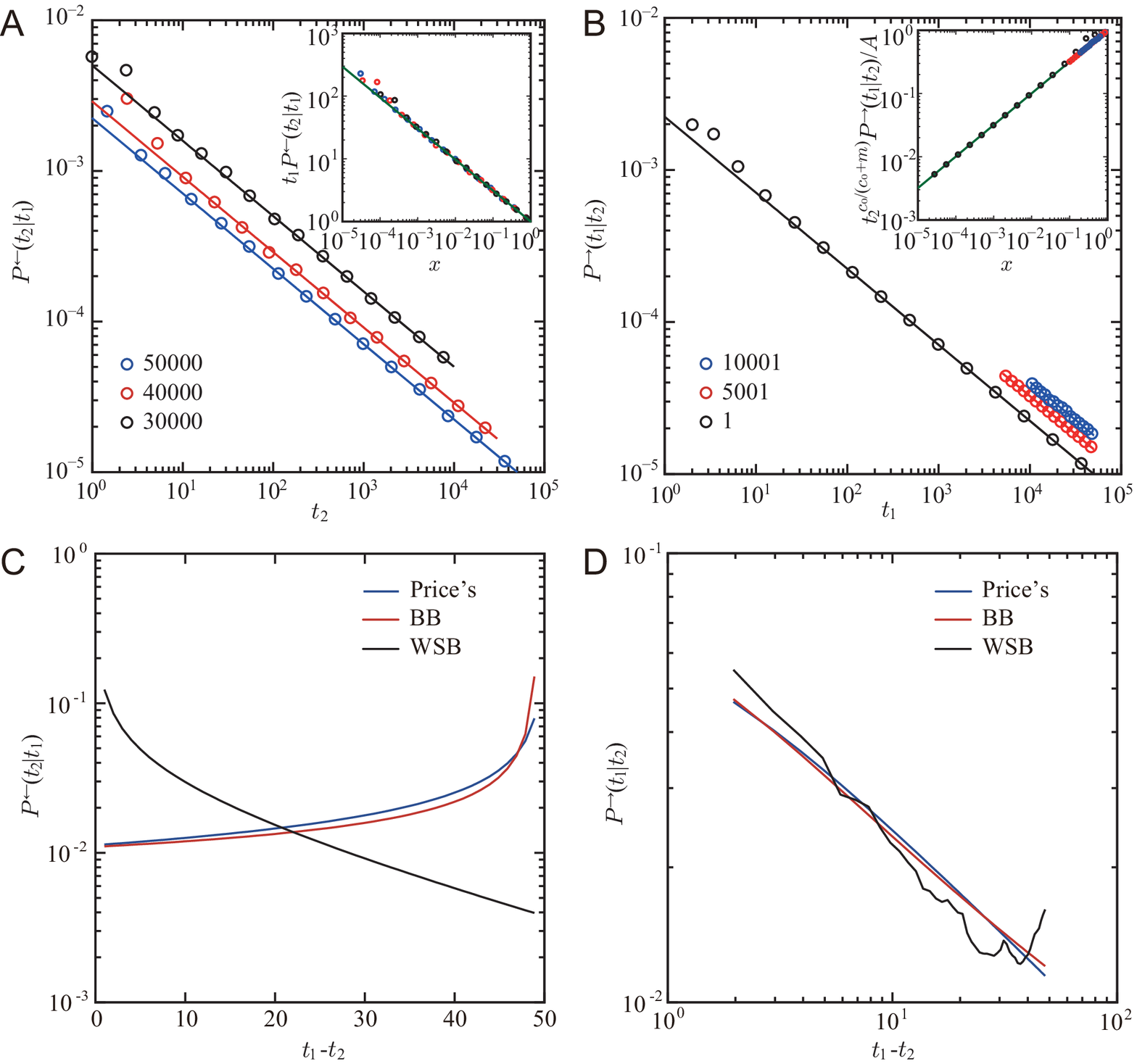}
\caption{{\bf Theoretical support for $P^{\leftarrow}(t)$ and $P^{\rightarrow}(t)$.}
{{\bf A} Price's model's prediction of $P^{\leftarrow}(t_2|t_1)$ for different time units. {Different circles are our simulations for different $t_1$}. Solid line shows our analytical calculation (\ref{eq:price_1}). Inset shows $P^{\leftarrow}(t_2|t_1)$ for different years collapse into a single curve, solid line showing prediction of (\ref{eq:price_3}). {\bf B} Price's model's prediction of $P^{\rightarrow}(t_1|t_2)$ for different years. {Different circles are our simulations for different $t_2$}. Solid line shows our analytical calculation (\ref{eq:price_2}). Inset shows $P^{\rightarrow}(t_1|t_2)$ for different years collapse into a single curve, solid line showing prediction of (\ref{eq:price_4}). {\bf C} $P^{\leftarrow}(t_2|t_1)$ predicted by Price's model, Bianconi-Barabasi model and Wang-Song-Barabasi model. {The WSB model} predicts the correct function over time consistent with our empirical results. {\bf D} $P^{\rightarrow}(t_1|t_2)$ predicted by Price's model, Bianconi-Barabasi model and Wang-Song-Barabasi model. Price's model and BB model predict approximately power law decay, while {the WSB model} is able to reproduce the upward tails given $r(t)$. }}
\label{model_simulation}
\end{figure}

What do existing models predict for \(P^{\leftarrow}(t)\) and \(P^{\rightarrow}(t)\)? To this end, we test three commonly used citation models with ascending levels of complexity: Price's model \citep{price1976general}, Bianconi-Barabasi model \citep{bianconi2001competition}, and Wang-Song-Barabasi model \citep{wang2013quantifying}.

Price's model assumes that each new paper cites $m$ references and the probability for an old paper to gain citations is proportional to its current citations plus a constant $c_0$ (Sec. S4.1). We calculate $P^{\leftarrow}(t)$ and $P^{\rightarrow}(t)$ analytically for this model, finding
\begin{equation}
P^{\leftarrow}(t_2|t_1)\sim\frac{1}{t_1}\left( 1- \frac{\Delta t}{t_1} \right)^{-\frac{m}{c_0+m}},
\label{eq:price_1}
\end{equation}
where $\Delta t=t_1-t_2$. (\ref{eq:price_1}) predicts that $P^{\leftarrow}(t_2|t_1)$ increases with $\Delta t$. That is, if we measure the citation age distribution following the retrospective approach at a given time $t_1$, the distribution increases as $\Delta t$ increases. This is due to the  cumulative advantage mechanism. Indeed, if we look further back in time, older papers are more likely to be characterized by a higher degree hence more likely to be cited. For prospective approach, we obtain 
\begin{equation}
P^{\rightarrow}(t_1|t_2)\sim A{\left(t_2+\Delta t\right)^{-c_0/(c_0+m)}}
\label{eq:price_2}
\end{equation}
where $A=\frac{1}{T^{m/(c_0+m)}-t_2^{m/(c_0+m)}}$.
Therefore the age distribution decays as a power law $\Delta t^{-c_0/(c_0+m)}$ as we look further in time. Interestingly, Price's model also predicts that $P^{\leftarrow}(t_2|t_1)$ collapse into a universal function for different $t_1$ after rescaling $x\equiv t_2/t_1$ (Fig. \ref{model_simulation}A):
\begin{equation}
P^{\leftarrow}(t_2|t_1)=\frac{c_0}{c_0+m}\frac{1}{t_1}x^{-m/(c_0+m)}
\label{eq:price_3}
\end{equation}
 $P^{\rightarrow}(t_1|t_2)$ can also be rescaled into a universal form as (Fig. \ref{model_simulation}B)
\begin{equation}
P^{\rightarrow}(t_1|t_2) = \frac{m}{c_0+m}At_2^{-c_0/(c_0+m)}x^{c_0/(c_0+m)}
\label{eq:price_4}
\end{equation}

The Bianconi-Barabasi (BB) model builds on Price's model by  adding a fitness parameter to each paper, \(\eta_i\) (Sec. S4.2). For retrospective approach, we find 
\begin{equation}
P^{\leftarrow}(t_2|t_1)\sim t_1\langle\eta x^{-\eta/C}\rangle,
\label{eq:bb1}
\end{equation} where $x\equiv t_2/t_1$, same as the rescaled time we previously defined for Price's model and $C$ is a normalization constant. Hence the BB model too predicts a retrospective distribution that increases with time. 
For prospective distribution, we find $P^{\rightarrow}(t_1|t_2)$ decays as
\begin{equation}
P^{\rightarrow}(t_1|t_2)\sim \langle\eta x^{1-\eta/C}/t_2\rangle/(\langle(T/t_2)^{\eta/C}\rangle-1). 
\label{eq:bb2}
\end{equation}
Eqs (\ref{eq:bb1}) and (\ref{eq:bb2}) also indicate that the shape of both distributions depends on the distribution of fitness. We calculate both distributions by assuming an exponential fitness distribution, i.e. $P(\eta) \sim e^{-\eta/\eta_0}$, a commonly used functional form \citep{kong2008experience}. 
We find that $P^{\rightarrow}(t_1|t_2)$ decays as a power law of $(\Delta t)^{-1}$, while $P^{\leftarrow}(t_2|t_1)$ again increases with time, albeit with a slower rate (Fig. \ref{model_simulation}CD).
Both distributions again collapse into universal forms if we use $x=t_2/t_1$:
\begin{equation}
P^{\leftarrow}(t_2|t_1) = \frac{1}{Ct_1\eta_0}\left(\frac{1}{\eta_0}+\frac{1}{C}\ln x\right)^{-2}
\end{equation}
\begin{equation}
P^{\rightarrow}(t_2|t_1) = \frac{\alpha_2x}{Ct_2(1-\alpha_2\eta_0)}\left(\frac{1}{\eta_0}+\frac{1}{C}\ln x\right)^{-2}
\end{equation}
where $\alpha_2 = 1/\eta_0-\ln(T/t_2)/C$. 

Finally we repeat our calculations for the Wang-Song-Barabasi (WSB) model \citep{wang2013quantifying}, which builds on the BB model but further incorporates an aging mechanism with the exponential growth of the system (Sec. S4.3). We find prospective distribution follows a lognormal distribution: 
\begin{equation}
P^{\rightarrow}(t_2|t_1)\sim \frac{1}{\sqrt{2\pi}\sigma\Delta t}e^{-(\ln(\Delta t)-\mu)^2/2\sigma^2},
\label{eq:wsb1}
\end{equation}
and the retrospective distribution follows lognormal distribution with exponential cutoff, i.e. 
\begin{equation}
P^{\leftarrow}(t_1|t_2)\sim \frac{1}{\sqrt{2\pi}\sigma\Delta t}e^{-(\ln(\Delta t)-\mu)^2/2\sigma^2-\beta\Delta t}.
\label{eq:wsb2}
\end{equation}
Somewhat surprisingly, these theoretical predictions are in remarkable agreement with our empirical results (Fig. \ref{model_simulation}CD). The accuracy of the WSB model is partly due to its ability to incorporate both aging and growth factors into the model. Together these results provide strong theoritical support for (\ref{eq:pretro}) and (\ref{eq:ppro}). 
\subsection{Connecting the Two Approaches}
The retrospective and prospective approaches are studied concomitantly in the literature, often pursued as separate lines of inquiry, which raise an interesting question: Are these two distributions connected?  Indeed, after all, they both measure the age of citations. Should we have measured the age distribution of \emph{all} citations,
both approaches would result in the same distribution, as if it did not matter whether we look retrospectively or prospectively in time. Yet despite the large amount of work on this topic (Table \ref{tab:age}), 
little is known if there exists any relationship between the two approaches.

\begin{figure}[h]
\centering
\includegraphics[width=1\linewidth]{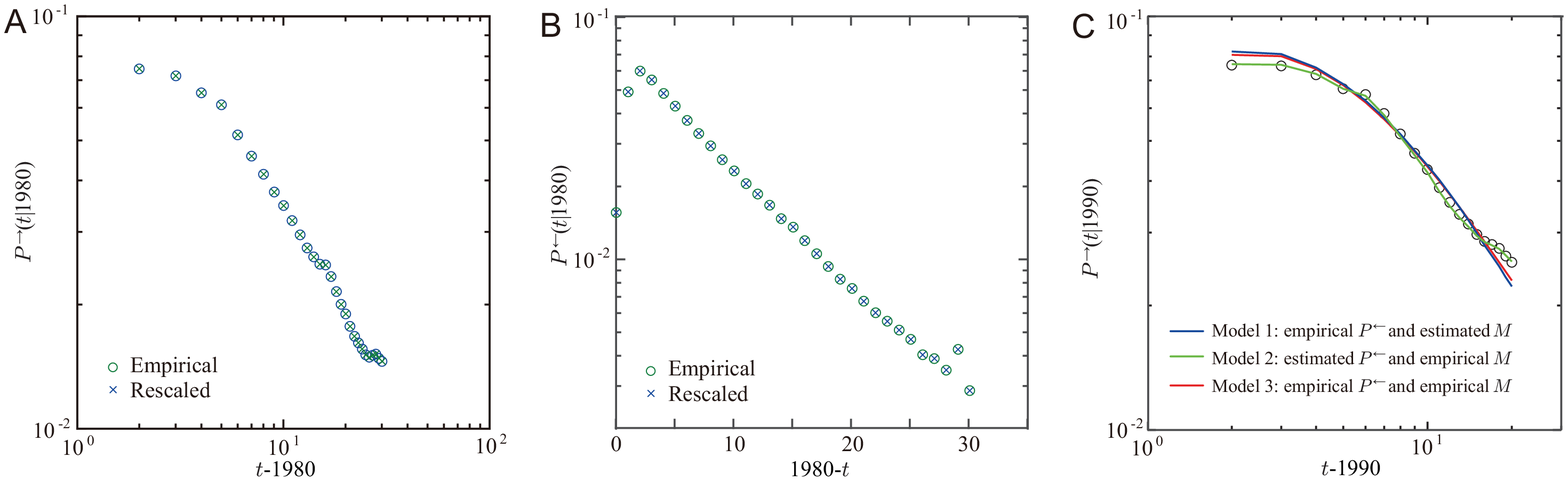}
\caption{{\bf Validations of (\ref{eq:convert0}) and (\ref{eq:convert1}).}
{ {\bf A} Using  \(P^{\leftarrow}(1980|t)\) data to calculate \(P^{\rightarrow}(t|1980)\) through rescaling formula (\ref{eq:convert0}). {\bf B}  Using \(P^{\rightarrow}(1980|t)\) data to calculate \(P^{\leftarrow}(t|1980)\) through rescaling formula (\ref{eq:convert1}). {\bf C} We use history data to estimate $M(t)$ and $P^{\leftarrow}(1990|t)$ and predict \(P^{\rightarrow}(t|1990)\) through rescaling formula (\ref{eq:convert0}).  Model 1 is based on estimated $M$ and empirical $P^{\leftarrow}$, Model 2 is based on empirical $M$ and estimated $P^{\leftarrow}$, while Model 3 is based on estimated $M$ and estimated $P^{\leftarrow}$. All three models show great predictive power.}}
\label{rescaling}
\end{figure}

To shed light on this question, we calculate $P(t_1, t_2)$, measuring the probability that a randomly selected citation connects a paper published in $t_1$ with a paper published in $t_2$. Because the difference between the two approaches lies in whether we fix the citing or cited papers, we can derive their citation age distribution from $P(t_1, t_2)$. {Next we show that expressing $P^{\leftarrow}$ or $P^{\rightarrow}$ in terms of $P(t_1, t_2)$ allows us to not only connect the two approaches, but also derive one approach from the other.} Indeed, $P(t_1,t_2)$ can be expressed as
 \begin{equation}
P(t_1,t_2)\sim P^{\leftarrow}(t_2|t_1)M(t_1)=P^{\rightarrow}(t_1|t_2)L(t_2)
\label{eq:rescale}
\end{equation}
where $L(t_2)=\int_{t_2}^{T}{P^{\leftarrow}(t_2|t_1)M(t_1)dt_1}$ measures the total number of citations papers published at $t_2$ receive. Therefore, (\ref{eq:rescale}) documents a precise mathematical connection between $P^{\rightarrow}(t_1|t_2)$ and $P^{\leftarrow}(t_2|t_1)$, mediated by $P(t_1,t_2)$, indicating one can obtain $P^{\rightarrow}(t)$ from $P^{\leftarrow}(t)$ or vice versa. 
Indeed, using (\ref{eq:rescale}) we obtain a rescaling formula for \(P^{\rightarrow}(t)\):
 \begin{equation}
P^{\rightarrow}(t_1|t_2)=\frac{P^{\leftarrow}(t_2|t_1)M(t_1)}{\int_{t_2}^{T}{P^{\leftarrow}(t_2|t_1)M(t_1)dt_1}},
\label{eq:convert0}
\end{equation}
which allows us to calculate the prospective distribution based solely on retrospective distributions (Sec. S5). 
Taking papers published in 1980 as an example, we collected \(P^{\leftarrow}(1980|t)\) for \(t>1980\) and rescaled them using (\ref{eq:convert0}). We then compare the rescaled results with  real  $P^{\rightarrow}(t)$, finding an excellent match between the two, reproducing not only the decay rate but also the upward tail (Fig. \ref{rescaling}A).

Similarly, we can solve for for \(P^{\leftarrow}(t)\) through proper rescaling of $P^{\rightarrow}(t)$ using (\ref{eq:rescale}) (Sec. S5). 
To this end we need to calculate $L(t)$ in terms of \(P^{\rightarrow}(t)\) from
 \begin{equation}
\int_{0}^{T}{P^{\rightarrow}(t_1|t_2)L(t_2)dt_2}=M(t_1).
\label{eq:fredholm}
\end{equation}
(\ref{eq:fredholm}) is the Fredholm integral equation of the first kind \citep{wazwaz2011}, which lacks analytical solution. 
We can, however, {consider a discretized version}, obtaining the rescaling formula for \(P^{\leftarrow}(t)\):
\begin{equation}
P^{\leftarrow}(t_2|t_1)=\frac{P^{\rightarrow}(t_1|t_2)[(\bm{P^{\rightarrow}})^{-1}\bm{M}](t_2)}{\bm{M}(t_1)}
\label{eq:convert1}
\end{equation}
where
\begin{align*}
&\bm{P^{\rightarrow}}=\left(P^{\rightarrow}(i|j)\Theta(i-j)\right)_{1\leq i,j\leq n}\\
&\bm{M}=[N(1)m(1)~~\cdots~~N(T)m(T)]^{T}
\end{align*}

To test its validity, we use (\ref{eq:convert1}) to numerically calculate the rescaled $P^{\rightarrow}$ , and compare it with real $P^{\leftarrow}$. 
More specifically, we take papers published in 1980, and measure all prospective distributions for every year between 1950 to 1980, 
i.e.~$P^{\rightarrow}(t_1|t_2)$ for $t_2$ ranging from 1950 to 1980. 
We then plug them into (\ref{eq:convert1}) to calculate the corresponding retrospective distributions. The obtained result yields excellent agreement with our empirical measurement (Fig. \ref{rescaling}B).

The remarkable agreement documented in Figure~\ref{rescaling}AB raises an interesting question: does our rescaling method have predictive power as well? To test this, we predict the prospective distribution in year 1990 by using retrospective distributions and the number of new citations prior to 1990. That is, we estimate values of \(M(t)\) and \(P^{\leftarrow}(t|t_0)\) for \(t=1990\) and \(t_0\) in \([1990,2010]\) in advance, seeking to obtain $P^{\rightarrow}(t|t_0)$ following (\ref{eq:convert0}).

Figure \ref{normalize}A indicate that \(M(t)\) is the product of $N(t)$, and $m(t)$. $P^{\leftarrow}(t|t_0)$, on the other hand, follows a lognormal distribution with exponential cutoff, whose parameters in (\ref{eq:pretro}) can be time-dependent. To test the feasibility of our approach, we impose a minimum set of assumptions to measure $M(t)$ and $P^{\leftarrow}(t|t_0)$, to check the degree to which the predicted $P^{\rightarrow}(t|t_0)$ matches the real distributions. More specifically, for \(M(t)\), we use an exponential function $\hat{M}(t)$ to fit the growth of references.  
We use a lognormal distribution with exponential cutoff to fit $P^{\leftarrow}(t|t_0)$.

To test the validity of our rescaling formula, we tried three models to predict $P^{\rightarrow}$ (Sec. S6.1). Model 1: we relax $M$ and predict $P^{\rightarrow}$ based on simulated $\hat{M}$ and real $P^{\leftarrow}$. Next, we build model 2 by relaxing $P^{\leftarrow}$ and hence using real $M$ and simulated $\hat{P}^{\leftarrow}$. Model 3 uses both simulated $\hat{M}$ and $\hat{P}^{\leftarrow}$ as approximations. Comparing with real data, we witness strong predictive power from all three models (Fig.~\ref{rescaling}C).

Our rescaling formula is also consistent with the derived functional form reported in preceding sections of this paper. Indeed, provided that $P^{\leftarrow}$ follows (\ref{eq:pretro}), (\ref{eq:convert0}) predicts that $P^{\rightarrow}$ follows the same functional form as (\ref{eq:ppro}), which is consistent with our empirical measurement (Sec. S6.2).

\subsection{Redner's Ansatz}
\begin{figure}[h]
\centering
\includegraphics[width=1\linewidth]{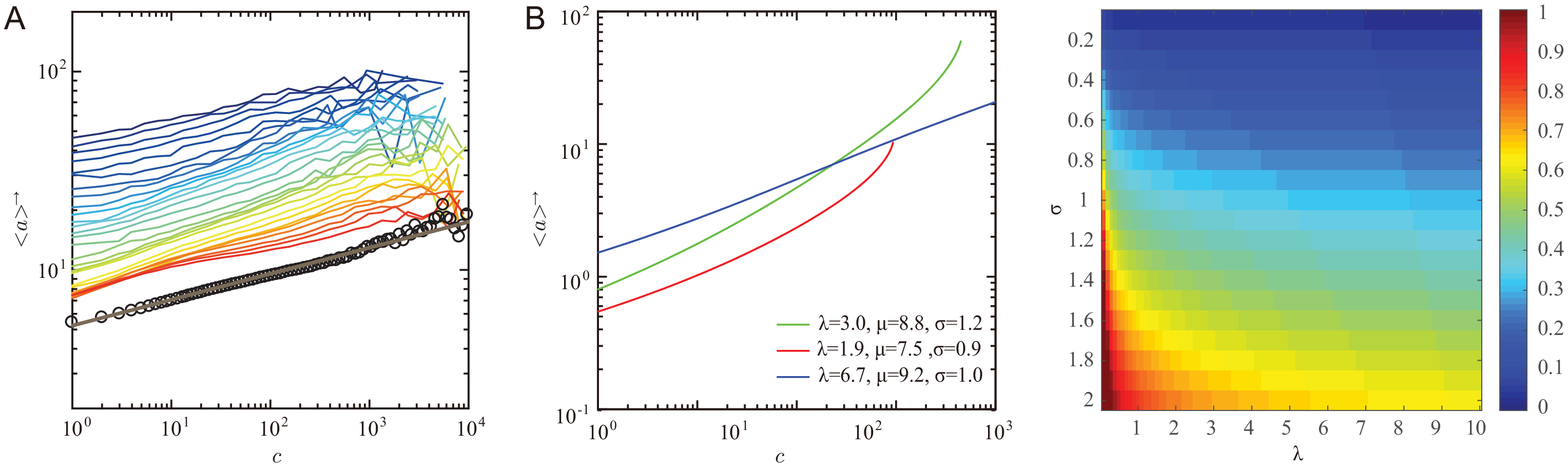}
\caption{{\bf Redner's ansatz.}{ {\bf A} Average age growth with citations of all papers on the WoS dataset (black circle) can be well fitted by power law (grey line, exponent is 0.13), we also see similar scaling when papers are grouped according to their ages (colored lines). Blue lines correspond to earlier years (1900-1902), while red lines correspond to later years (1978-1980). We also observe a decrease of average citation age for recent years since these papers can only be cited by more recent papers. {\bf B} Simulation of average citation growth for different \((\lambda,\mu,\sigma)\) using Wang-Song-Barabasi model, the power law scaling is observed in a long range.} {\bf C} 
Varying $\lambda$ and $\sigma$ parameters in the WSB model can result in a wide range of scaling exponent.
Color indicates the scaling exponent measured as slopes in log-log plots similar to A and B.}
\label{age}
\end{figure}

How does the citation aging structure of a paper connect to its citations? 
To answer this question, {we consider an interesting ansatz} raised by Redner \citep{redner2005}: 
there appears to exist a power law relationship between average citation age and total citations a paper receives. Redner took all papers with 500 or fewer citations published in the American Physics Society (APS) dataset, and found that the average age of all citations to a paper, i.e. $\langle a\rangle^{\rightarrow}$, grows with the number of citations its receives, \(c\), following \(\langle a\rangle^{\rightarrow}\sim c^{\alpha}\). We repeated this analysis on the WoS dataset, finding $\langle a\rangle^{\rightarrow}$ and $c$ indeed follow a power law scaling ($\alpha\approx0.13$) in the small to medium $c$ regime ($c<500$) (Fig. \ref{age}A), thus independently confirming Redner's ansatz. We also find {significant deviations} in the large $c$ regime ($c>500$), where $\langle a\rangle^{\rightarrow}$ appears to increase with $c$ at a faster rate than power law.

To test the robustness of this ansatz, we grouped {papers based on their publication times and measure the average citation age for papers with different citations within each group}, finding again a power law scaling between $c$ and $\langle a\rangle^{\rightarrow}$  (Fig. \ref{age}A). Next, we show that this power law scaling can be explained by the inherent citation dynamics individual papers follow. {To approximate the citation dynamics of a paper}, we use the WSB model to generate citations of a paper over time and measure their relationship with the average citation age (Sec. S7). 
As we vary its $(\lambda,\mu,\sigma)$ parameters, we find that across a wide range of $c$ values, $\langle a\rangle^{\rightarrow}$ can be approximated by power law relationships. The model also successfully captures the acceleration phase of the relationship in large $c$ region (Fig. \ref{age}B). The scaling exponent of the power law relationship $\alpha$ depends on the specific $(\lambda,\mu,\sigma)$ parameters, hence varying $(\lambda,\mu,\sigma)$ through the parameter space allows us to reproduce the scaling behavior with a wide range of exponents (Fig. \ref{age}C). 

\section{Discussion}

{Understanding the temporal behavior of citations is of fundamental importance. In this paper, we carried out a large-scale quantitative analysis on how time affects citations. We began by reviewing and comparing the most important papers on this topic, finding significant inconsistency among prior studies in terms of both their measurement approaches and conclusions. To systematically measure citation age distributions, we used the Web of Science dataset to test different functional forms through maximum likelihood estimation. We found neither of the two prevailing approaches taken by current studies can be well captured by existing models. 
Moreover, we found ubiquitous upward tails characterizing $P^{\rightarrow}$, indicating that citation age distributions are affected by growth of papers and citations. To account for such impact, we further defined and calculated normalized distributions for both approaches. We found both normalized distributions can be well captured by lognormal functions, which not only allows us to fit age distributions more accurately, but also explains upward tails observed in empirical data. }

{Given that all the distributions reported in Table \ref{tab:age} are inherently quite similar (Fig. \ref{overview}), here instead of focusing on visually how distributions fit the data, we aimed to look at this question through a theoretical lens. Indeed, we situated the observations within a theoretical framework connecting both distributions through mathematical relationships and network models, allowing us to rescale both distributions and reason about the most appropriate functional forms. Indeed, the main virtue of this paper lies in the theoretical framework that helps us connect the retrospective and prospective approaches.} We validated the newly derived distributions both empirically and theoretically. 
Through theoretical and empirical validations, we show that either distribution can be mathematically obtained from another distribution because of deeper underlying connections between the two approaches.

We then placed our results in the context of citation models, 
analytically calculating the citation age distributions for both approaches in three citation model systems. 
Compared with our empirical results, we find the Wang-Song-Barabasi model provides an accurate prediction for the functional forms of citation age distributions we obtained, offering theoretical support for our empirical findings. 
On the other hand, our results also provide additional evidence supporting the accuracy of the WSB model in characterizing the temporal behavior of citations. We concluded the paper with a discussion on the Redner's ansatz that postulates a power law scaling between citations and average citation ages, finding it again can be well captured by the WSB model.

{Taken together, this paper provides systematic answer to the mathematical form of citation age distributions. We believe the results reported in this paper provide a new empirical and theoretical basis for future studies on temporal analyses of citations, and will play an increasingly important role as our quantitative understanding of science deepens. Future directions include extending our analyses to examing their dependence on contextual factors, such as disciplines, countries or publishing venues, remains unknown, which could offer a more holistic understanding of the time dimension of science.}

\section{ACKNOWLEDGMENTS}
The authors wish to thank Zhen Lei, Chaoming Song, and A.-L. Barab{\'a}si for helpful discussions. This work is supported by the Air Force Office of Scientific Research under award number FA9550-15-1-0162 and FA9550-17-1-0089.

\bibliography{science}

\end{document}